\documentclass[%
reprint,
superscriptaddress,
preprintnumbers,
amsmath,amssymb,
aps,
prd,
]{revtex4-2}

\usepackage{float}
\usepackage{graphicx}
\usepackage{dcolumn}
\usepackage{bm}
\usepackage{bbold}
\usepackage{braket}
\usepackage{amssymb,amsmath}
\usepackage{bbm}
\usepackage{mathrsfs}

\usepackage{color}
\usepackage{amsfonts}
\usepackage{subfigure}
\usepackage{array}
\usepackage{enumerate}

\newcolumntype{L}{>{\centering\arraybackslash}m{3cm}}

\definecolor{blue}{rgb}{0,0,1}

\definecolor{green}{rgb}{0,1,0}

\definecolor{red}{rgb}{1,0,0}

\definecolor{gray}{rgb}{.5,.5,.5}

\definecolor{darkgreen}{rgb}{.0,.5,.0}

\usepackage{xspace}
\usepackage{siunitx}
\usepackage{xfrac}
\usepackage{hyperref}
\usepackage[nameinlink]{cleveref}
\usepackage{appendix}
\usepackage{slashed}
\usepackage{xifthen}
\usepackage{xcolor}
\hypersetup{
	colorlinks,
	linkcolor={red!75!black},
	citecolor={blue!75!black},
	urlcolor={blue!75!black}
}
\usepackage{changes}

\def\eqref#1{\Cref{#1}}

\def\app#1{\hyperref[#1]{App.~\ref{#1}}}
\def\app#1{\Cref{#1}}

\def\0#1#2{\frac{#1}{#2}}

\graphicspath{{./figures/}{./}}

\begin{document}
	
\title{Functional renormalization group study of anomalous magnetic moment in a low energy effective theory}

\author{Rui Wen}
\email{rwen@ucas.ac.cn}
\affiliation{School of Nuclear Science and Technology, University of Chinese Academy of Sciences, Beijing, 100049,  P.R. China}

\author{Chuang Huang}
\email{huang@thphys.uni-heidelberg.de}
\affiliation{Institut f{\"u}r Theoretische Physik, Universit{\"a}t Heidelberg, Philosophenweg 16, 69120 Heidelberg, Germany}

\author{Mei Huang}
\email{huangmei@ucas.ac.cn}
\affiliation{School of Nuclear Science and Technology, University of Chinese Academy of Sciences, Beijing, 100049,  P.R. China}

\begin{abstract}

The quark anomalous magnetic moments (AMMs) are investigated in a 2-flavor low-energy effective theory within the functional renormalization group (FRG) approach under an external magnetic field. The Schwinger formalism is adopted for quark propagators, and Fierz-complete four-quark scatterings are self-consistently included through the renormalization group flows. We find that the quark AMMs are dynamically generated with the chiral symmetry breaking, and the magnitude of the AMM of the down quark is around 4 times larger than that of the up quark. The transverse AMMs and the longitudinal d-quark AMM monotonically decrease with the magnetic field strength, while the longitudinal u-quark AMM slightly increases with the magnetic field strength. At $B=0$, the magnetic moments of proton and neutron are computed using the constituent quark model, which are close to the experimental values.
\end{abstract}


\maketitle
	
\section{Introduction}
\label{sec:Introduction}

Strong-interaction matter in an external magnetic field has attracted much attention in recent years. It can be created through non-central heavy-ion collisions and can exist in magnetars and the early universe \cite{Skokov:2009qp, Deng:2012pc, Vachaspati:1991nm, Durrer:2013pga, Kiuchi:2015sga}. Non-central collisions of two high-speed nuclei moving in opposite directions can create strong magnetic fields of up to order $\sim 10^{18}$ Gauss \cite{Skokov:2009qp, Deng:2012pc}. Magnetic fields on the surfaces of magnetars are estimated to be in the range of $\sim 10^{10}-10^{15}$ Gauss \cite{haensel2007, Duncan:1992hi}, and may reach up to $\sim 10^{18}$ Gauss in their interiors \cite{1991ApJ...383..745L}. Such strong magnetic fields influence the chiral and deconfinement phase transitions and give rise to various novel observable phenomena, such as the chiral magnetic effect \cite{Kharzeev:2007jp, Kharzeev:2010gr}, magnetic catalysis and inverse magnetic catalysis \cite{Fukushima:2012xw, Klevansky:1989vi, Klimenko:1990rh, Gusynin:1995nb, Bali:2012zg, Tomiya:2019nym, Andersen:2021lnk, Bali:2011qj}, the non-monotonic behavior of the charged pion mass \cite{Ding:2020hxw}, and diamagnetism around the critical temperature \cite{Bali:2014kia}. Many investigations have been carried out from Lattice QCD simulations \cite{Bali:2011qj, Bali:2012jv, Bali:2014kia, Bali:2017ian, Bignell:2020dze, Bornyakov:2013eya, Ding:2020hxw, Ding:2022tqn,Ding:2025siy,Ding:2025jfz}; effective model studies, such as the Nambu-Jona-Lasinio (NJL) model \cite{Inagaki:2003yi, Chao:2014wla, Yu:2014xoa, Avancini:2016fgq, Coppola:2018vkw, Coppola:2019uyr, Mei:2022dkd, Mei:2024rjg} and the quark-meson model \cite{Wen:2023qcz, Kamikado:2013pya, Kamikado:2014bua, Ayala:2018zat}; and theoretical approaches such as renormalization group \cite{Fukushima:2012xw, Braun:2014fua, Fu:2017vvg} and Dyson-Schwinger equations \cite{Mueller:2015fka}. For recent reviews, see \cite{Hattori:2023egw, Adhikari:2024bfa}.

The anomalous magnetic moment (AMM), which is already discussed in standard textbooks on QED \cite{Peskin:1995ev, weinberg2002quantum}, can also be dynamically generated for quarks through dynamical chiral symmetry breaking in QCD \cite{Chang:2010hb}. Recent studies have shown that the quark AMM influences the QCD phase diagram and mesonic properties in the presence of external magnetic fields \cite{Fayazbakhsh:2014mca, Chaudhuri:2019lbw, Xu:2020yag, Yang:2025zcs, Mao:2022dqn, Strickland:2012vu, Mondal:2024eaw, Farias:2021fci, Chao:2022bbv, Mei:2020jzn}, providing a possible explanation for the inverse magnetic catalysis puzzle. However, most of these studies treated the quark AMMs as free parameters, and some inappropriate parameter choices and regularization led to unphysical results. Therefore, a self-consistent calculation of the quark AMM is valuable. In \cite{Lin:2021bqv}, the AMM is calculated from the fermion-photon vertex using the Schwinger proper-time formalism in a low magnetic field. In \cite{Ghosh:2021dlo}, the authors employ the two-flavor NJL model in the mean field approximation, calculate the photon-quark-antiquark vertex, and obtain the quark AMM under external magnetic fields. In \cite{Fraga:2024klm}, the authors calculate the quark AMM using perturbative QCD with the lowest-Landau level approximation and effective gluon mass.

In this work, we employ a 2-flavor low-energy effective theory for calculating the quark mass and anomalous magnetic moments under a magnetic field within the functional renormalization group (FRG) approach. FRG is a functional realization of Wilson's RG ideas \cite{Wetterich:1992yh, Pawlowski:2005xe}. It serves as a powerful tool for calculations in non-perturbative theories. The Schwinger formalism of quark propagators under a magnetic field is employed, and Fierz-complete four-quark scatterings are self-consistently included through the RG flows. The up and down quark AMMs in the directions parallel and perpendicular to the magnetic field are calculated, which could serve as inputs for other model studies.

This paper is organized as follows. In \cref{sec:2-flavorLEFT}, we introduce the 2-flavor low-energy effective theory under a magnetic field within the FRG approach. The calculation of the anomalous magnetic moment is presented in \cref{sec:AMM}.  In \cref{sec:results}, we show the numerical results in our calculations, including the quark masses, four-quark couplings and  anomalous magnetic moments as functions of the strength of magnetic field. The summary and outlook are given in \cref{sec:summary}.

\section{2-flavor low energy effective theory}
\label{sec:2-flavorLEFT}

In this work, we adopt a 2-flavor low-energy effective theory with the kinetic term of the quarks and the four-quark scattering terms\cite{Nambu:1961fr, Fu:2022uow}. The effective action in Euclidean space reads:
\begin{align}\label{eq:effectiveaction}
\Gamma_k[\bar{q},q]=&\int_x  \bar{q} (\gamma_\mu D_\mu+m) q +\sum_\alpha \lambda^{(\alpha)}_{4q} \mathcal{T}^{(\alpha)}_{4q,ijlm} \bar{q}_i q_j \bar{q}_l q_m 
\end{align}
with a shorthand notation $\int_x= \int d^4 x$. Here $D_\mu \equiv \partial_\mu - i Q A_{\mu,0} $ and $Q$ denotes the quark electric charge. For 2-flavor case, $Q=\text{diag}(2/3,-1/3)e$ with $q = (u,d)^T$. $A_{\mu,0}$ denotes a background homogeneous magnetic field. Without loss of generality, we assume a magnetic field along the $z$-direction and adopted the Landau gauge, i.e., $A_{\mu,0}=(0,0,xB,0)$.

The second term of \cref{eq:effectiveaction} denotes the four-quark interaction $\Gamma_{4q}$ with couplings $\lambda^{(\alpha)}_{4q}$ and tensor structures $\mathcal{T}^{(\alpha)}_{4q,ijlm}$. In this work, we consider the isospin-symmetric Fierz-complete tensor structures and ignore the momentum dependence, which are listed in \cite{Fu:2022uow, Fu:2024ysj}:
\begin{align}
\alpha \in \{  &\pi,\sigma, \eta, a, (V\pm A), (V-A)^{adj},\nonumber \\ 
&(S\pm P)_{-}^{adj},(S+P)_{+}^{adj}\}.
\end{align}
$\mathcal{T}^{(\alpha)}_{4q,ijlm}$ are antisymmetric under the commutation of pairs of indices $i,j$ and $l,m$, i.e. 
\begin{align}
\mathcal{T}^{(\alpha)}_{4q,ijlm} =- \mathcal{T}^{(\alpha)}_{4q,lmij}.
\end{align}
There is no doubt that the magnetic field breaks isospin symmetry, which leads to splitting into 20 tensor structures. These will be considered in future work.

The quark propagator in a homogeneous magnetic field is $\hat G=\text{diag}[G(x,y;q_u),G(x,y;q_d)]$, with the Schwinger formalism of quark propagators \cite{Schwinger:1951nm}
\begin{align}
G(x,y;q_f)=\Phi(x,y)\int\frac{d^4p}{(2 \pi)^4} e^{-i p (x-y)} \tilde{G}(p;q_f).
\end{align}
Here $\Phi(x,y)=q_f B(x^1+y^1)(x^2-y^2)/2$ is the Schwinger phase factor and $\tilde{G}(p;q_f)$ is the translation invariance part. In Euclidean space, the translation invariance part $\tilde{G}(p;q_f)$ reads
\begin{align}
\tilde{G}&(p;q_f)= \int_0^\infty ds \exp\bigg[- s\bigg((1+r_{f,s})^2 \bigg(p_\parallel^2 \nonumber \\
&+p_\perp^2\frac{\tanh(q_f B s)}{q_f B s}\bigg) + m^2\bigg) \bigg] \nonumber \\
& \cdot [(-i p_\parallel \gamma_\parallel (1+r_{f,s}) + m) (1+i\gamma_1 \gamma_2 \tanh(q_f B s))\nonumber \\
&   -i   p_\perp \gamma_\perp(1+r_{f,s}) (1 - \tanh^2(q_f B s))] .
\end{align}
Here, $r_{f,s}$ denotes the magnetic dependent 4d regulator of quarks. In this work, we use a simple exponential regulator
\begin{align}
r_{f,s}=\frac{e^{-x}}{x} , \quad  \quad x=\frac{p_\parallel^2+p_\perp^2\tanh(q_f B s)/(q_f B s)}{k^2}.
\end{align}
Here $k$ denotes the infrared (IR) cutoff scale, i.e., the RG scale. The regulator $r_{f,s}$ temporarily suppresses quantum fluctuations of momenta below the scale $k$ and is eventually removed as $k \rightarrow 0$. This regulator is a common choice for momentum-dependent approximations. We also employ the Litim regulator and get qualitatively the same results. For more discussion about the regulator dependence, see \cite{Fu:2022uow}.

\begin{figure}[t]
\includegraphics[width=0.49\textwidth]{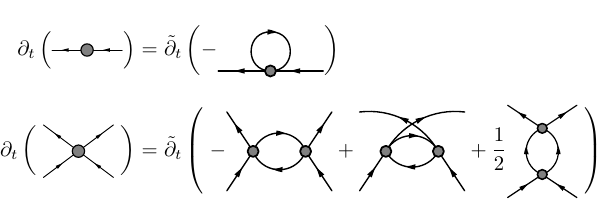}
\caption{Feynman diagrams of flow equations of two- and four-point quark correlation functions, see also \cite{Fu:2022uow}.}\label{quark24_eq}
\end{figure}

The flow equations for the two- and four-point quark correlation functions are derived from the Wetterich equation \cite{Wetterich:1992yh}, which are depicted in \Cref{quark24_eq}. The flow equation for the quark mass is obtained by projecting the flow of the two-point correlation function onto the scalar channel:
\begin{align}
\partial_t m = \int\frac{d q^4}{(2\pi)^4}\frac{1}{4N_c N_f} \text{Tr}[\Gamma_{4q} \tilde{\partial}_t \hat{G}].
\end{align}
Here, the RG time $t\equiv \ln (\Lambda/k)$, and the partial derivative with a tilde, $\tilde{\partial}_t$, denotes that it acts only on the RG scale through the regulators in propagators. Traces are taken over color, flavor, and Dirac space. The detailed expression of the flow equation for the quark mass is
\begin{align}
\partial_t m &= \frac{1}{6} (9 \lambda^a-3 \lambda^\eta-9 \lambda^\pi  -69 \lambda^\sigma-16 \lambda^{(S+P)_{-}^{adj}} \nonumber \\
&+32 \lambda^{(S+P)_{+}^{adj}}+24 \lambda^{(V+A)}) (\mathscr{L}_m(u)+\mathscr{L}_m(d)),
\end{align}
with the loop function for quark mass
\begin{align}
\mathscr{L}_m(f)\equiv & \int \frac{d^4p}{(2 \pi)^4}\int_0^{\infty} ds \mathscr{E}(f,s) m \partial_t r_{f,s}(r_{f,s}+1)\nonumber \\
 &\left(p_{\perp}^2\frac{\tanh(q_fB s)}{q_fB}+p_{\parallel}^2 s\right).
\end{align}
Here, for convenience, we define a shorthand notation
\begin{align}
&\mathscr{E}(f,s) \equiv \nonumber \\
&\exp\bigg[- s\bigg((1+r_{f,s})^2 \bigg(p_\parallel^2 +p_\perp^2\frac{\tanh(q_f B s)}{q_f B s}\bigg) + m^2\bigg) \bigg].
\end{align}
The flow equations for the four-quark interaction couplings are calculated in a manner similar to \cite{Fu:2022uow}, with the aid of the \texttt{FormTracer} package \cite{Cyrol:2016zqb}. The calculation is straightforward but tedious, and the expressions are provided in \cite{NJLtypeFRG}.

Notably, the Schwinger phase factor cancels out in the calculation of the two-point correlation function. For the four-point correlation function, the Schwinger phase factor cancels out in same-flavor quark loops but remains in the calculation of u-d mixing quark loop. In this work, we neglect the contribution of the Schwinger phase in the latter case and leave its treatment for future work. For further discussion on the Schwinger phase, see, e.g., \cite{Coppola:2018vkw, Mao:2018dqe, GomezDumm:2023owj}. 

\section{Anomalous magnetic moment}
\label{sec:AMM}

\begin{figure}[t]
\includegraphics[width=0.45\textwidth]{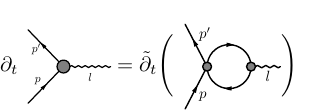}
\caption{Feynman diagram of the flow equation of the quark-photon vertex.}\label{quarkphoton-equ}
\end{figure}

The effective action of the full quark-photon vertex, including non-classical tensor structures in the Landau gauge, is
\begin{align}
\Gamma_{A\bar q q,\mu}&=\delta (p+p'+l)Q\otimes\sum_{i=1}^{8} \Gamma_{A\bar q q,\mu}^{(i)} \,, \\
\Gamma_{A\bar q q,\mu}^{(i)}&= \lambda^{(i)}_{A\bar{q} q} [\mathcal{T}^{(i)}]_\mu(p,p',l)\,.
\end{align}
Here, $p, p', l$ are the external momenta of quark, anti-quark, and photon, respectively. $\mathcal{T}^{(i)}$ denote the tensor structures listed in \cite{Ihssen:2024miv, Fu:2025hcm}. The classical tensor structure is
\begin{align}
 [\mathcal{T}^{(1)}]_\mu (p, p', l)=&\Pi_{\mu\nu}^\perp(l) i \gamma_\nu,
 \end{align}
with the transverse projection operator $\Pi_{\mu\nu}^\perp(l)=\delta_{\mu\nu}-\frac{l_\mu l_\nu}{l^2}$. The anomalous magnetic moment is related to the tensor structure
\begin{align}
[\mathcal{T}^{(4)}]_\mu (p, p', l) =&  i \sigma_{\mu\nu} (l)_\nu.
\end{align}
By comparing with the quark-photon effective vertex given in \cite{Ghosh:2021dlo}, we define the dimensionless form factors:
\begin{align}
F_{2,f}(p,p',l)\equiv2 m \lambda^{(4)}_{A\bar{q}_f q_f}(p,p',l)\,, \quad f\in\{u,d\},
\end{align}
and the quarks AMM:
\begin{align}
\kappa_f\equiv\lim_{l\rightarrow 0}\lambda^{(4)}_{A\bar{q}_f q_f}(p,p',l)\,, \quad f\in\{u,d\},
\end{align}

The flow equations of the quark-photon vertex are
\begin{align}\label{eq:quark-photon}
\partial_t \Gamma_{A\bar{q}q,\mu}=\int \frac{d q^4}{(2\pi)^4}\tilde{\partial}_t ( \Gamma_{A\bar q q,\mu}\hat{G} \Gamma_{4q} \hat{G}),
\end{align}
which are depicted as \Cref{quarkphoton-equ}. Notably, in this work we adopt the momentum-independent approximation for the four-quark interaction. As a result, the quark-photon vertex only depends on the external momentum of the photon, $l$, and we abbreviate the form factors as $F_{2,f}(l)$. We project the flow equations onto $\mathcal{T}^{(4)}$ tensor structure
\begin{align}
\partial_t F_{2,f}(l)=\frac{-2im}{12 N_c l^2 q_f} \text{Tr}_{d,c}[l_\nu \sigma_{\mu\nu} \partial_t \Gamma_{A\bar{q}_fq_f,\mu} ],
\end{align}
and
\begin{align}
\partial_t \kappa_f=\lim_{l\rightarrow 0}\frac{-i}{12 N_c l^2 q_f} \text{Tr}_{d,c}[l_\nu \sigma_{\mu\nu} \partial_t \Gamma_{A\bar{q}_fq_f,\mu} ].
\end{align}
Here, the indices $d$ and $c$ indicate that the traces are taken only over the Dirac and color space, respectively. Since the classical quark-photon vertex is much larger than the other tensor structures, we adopt the approximation $\Gamma_{A\bar q q,\mu} \sim S_{A\bar q q,\mu} = i \Pi_{\mu\nu}^\perp(l) \gamma_\nu$ on the right-hand side of \cref{eq:quark-photon}. The Schwinger phase factor cancels out in this diagram. With the aid of the \texttt{FormTracer} package \cite{Cyrol:2016zqb}, we obtain the flow equations for the u-quark AMMs
\begin{align}
&\partial_t \kappa_{u,\parallel/\perp}=-\frac{1}{6 q_u} \nonumber \\
&\bigg[ 2 (3 \lambda^a- 3 \lambda^\pi - 8 \lambda^{(S+P)_-^{adj}} + 8 \lambda^{(S+P)_+^{adj}})q_d\mathscr{L}_{\kappa,\parallel/\perp}(d)\nonumber\\
&+(3\lambda^a - 3\lambda^\eta - 3\lambda^\pi + 3\lambda^\sigma + 16\lambda^{(S+P)_+^{adj}})q_u\mathscr{L}_{\kappa,\parallel/\perp}(u) \bigg].
\end{align}
Here, the loop functions are defined as
\begin{align}\label{eq:loop_kappa1}
&\mathscr{L}_{\kappa,\parallel}(f)=\int \frac{d ^4 q}{(2 \pi)^4} \int_0^\infty ds\int_0^\infty dr \mathscr{E}(f,r) \mathscr{E}(f,s) m\nonumber \\
&[\partial_t r_{f,r} (1  - 2 (1 + r_{f,r})^2) ( q_\parallel^2 r+ q_\perp^2 \tanh(q_f Br)/(q_fB)) \nonumber \\
&- 2 \partial_t r_{f,s} (1 + r_{f,r}) (1 + r_{f,s}) (q_\parallel^2 s+ q_\perp^2 \tanh(q_f Bs)/(q_fB)]\nonumber \\
& (-1 + \tanh(q_f Br)  \tanh(q_f B s)/3),
\end{align}
and
\begin{align}\label{eq:loop_kappa2}
&\mathscr{L}_{\kappa,\perp}(f)=\int \frac{d ^4 q}{(2 \pi)^4} \int_0^\infty ds\int_0^\infty dr \mathscr{E}(f,r) \mathscr{E}(f,s) m\nonumber \\
&[\partial_t r_{f,r} (1  - 2 (1 + r_{f,r})^2) ( q_\parallel^2 r+ q_\perp^2 \tanh(q_f Br)/(q_fB)) \nonumber \\
&- 2 \partial_t r_{f,s} (1 + r_{f,r}) (1 + r_{f,s}) (q_\parallel^2 s+ q_\perp^2 \tanh(q_f Bs)/(q_fB)]\nonumber \\
&(-1 + \tanh^2(q_uBr)).
\end{align}
For the expression of the d-quark, change $u \leftrightarrow d$, and $\partial_t F_{2,f}(0)=2 m \partial_t \kappa_f$. “Parallel” (longitudinal) or “perpendicular” (transverse) refers to whether the direction in which the photon's external momentum approaches zero is parallel or perpendicular to the external magnetic field.

\begin{figure*}[t]
\includegraphics[width=1\textwidth]{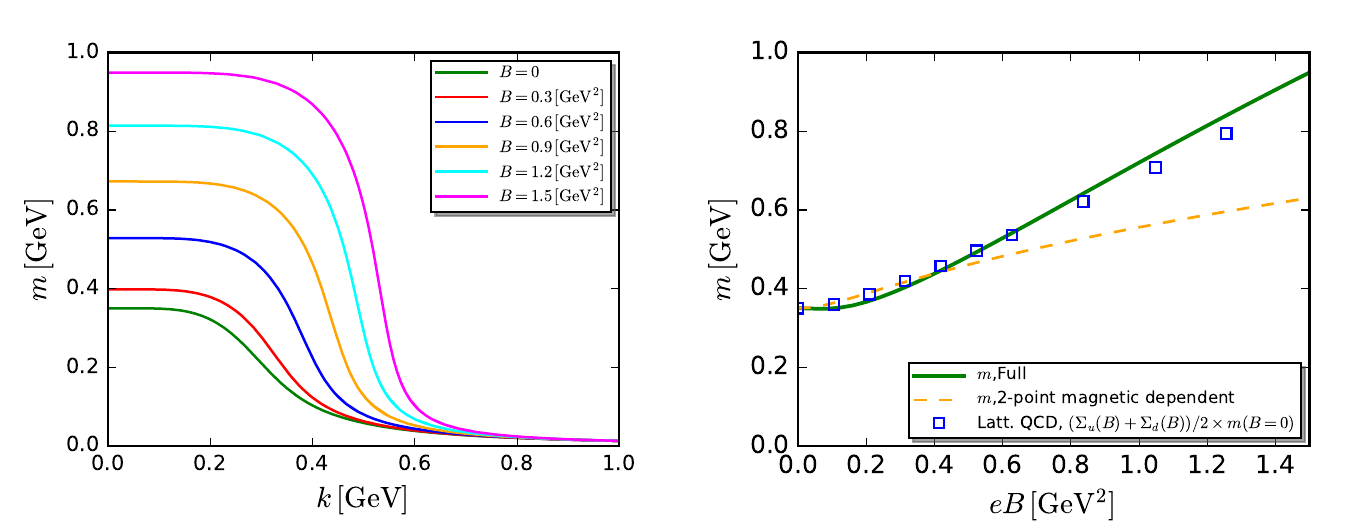}
\caption{Left panel: The quark masses as functions of RG scale $k$ with several values of the strength of the magnetic field. Right panel: The quark masses as functions of the strength of magnetic field. }\label{fig:mf}
\end{figure*}

\begin{figure*}[t]
\includegraphics[width=1\textwidth]{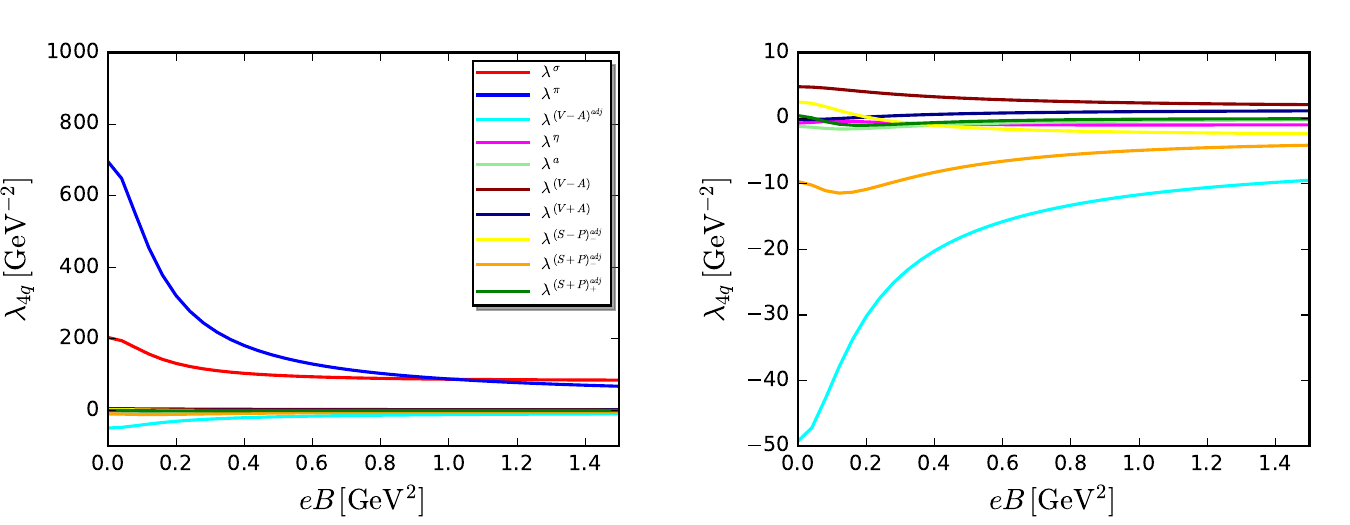}
\caption{Left panel: The four-quark couplings of Fierz complete channels as functions of the strength of the magnetic field. Right panel: Non-dominant four-quark couplings of the Fierz complete channels, which is the zoomed-in version of the left panel.}\label{fig:lam}
\end{figure*}

\begin{figure}[t]
\includegraphics[width=0.5\textwidth]{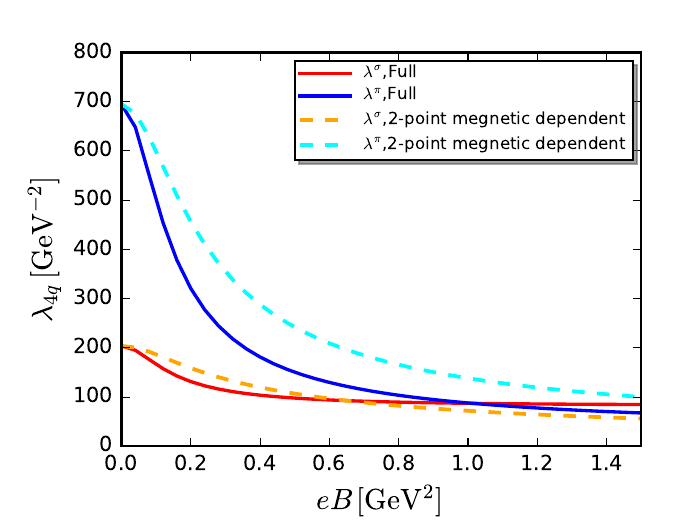}
\caption{Comparison of the dominant four-quark couplings as functions of the strength of the magnetic field between full calculation and 2-point magnetic dependent truncation. }\label{fig:lamcompare}
\end{figure}

Notably, if we ignore the non-dominant channels, i.e., $\alpha\notin\{\pi,\sigma\}$, and the splitting between the scalar and pseudoscalar couplings, the u-quark AMM depends only on the d-quark loop contribution, and vice versa. Based on this, we obtain $\kappa_d/\kappa_u=(q_u^2/q_d^2) \cdot (\mathscr{L}_{\kappa}(u)/ \mathscr{L}_{\kappa}(d))=4 \mathscr{L}_{\kappa}(u)/ \mathscr{L}_{\kappa}(d)$, which are consistent with \cite{Ghosh:2021dlo}.

The magnetic moments of quarks are connected to the form factors, which are given as \cite{Fayazbakhsh:2014mca, Ghosh:2021dlo}:
\begin{align}\label{eq:magneticmoments1}
\mu_f=q_f (1+F_{2,f}(0)) \frac{m_N}{m} \mu_N\,,\quad f\in\{u,d\},
\end{align}
where $m_N=938$ MeV is the nucleon mass and $\mu_N=e/(2 m_N)$ denotes the nuclear magneton. In the constituent quark model, the proton and neutron magnetic moments are given as
\begin{align}
\mu_\text{proton}&=\frac{1}{3} (4 \mu_u -\mu_d),  \label{eq:magneticmoments2}\\
\mu_\text{neutron}&=\frac{1}{3} (4 \mu_d -\mu_u).\label{eq:magneticmoments3}
\end{align}

\section{numerical results}
\label{sec:results}

\begin{figure}[t]
\includegraphics[width=0.5\textwidth]{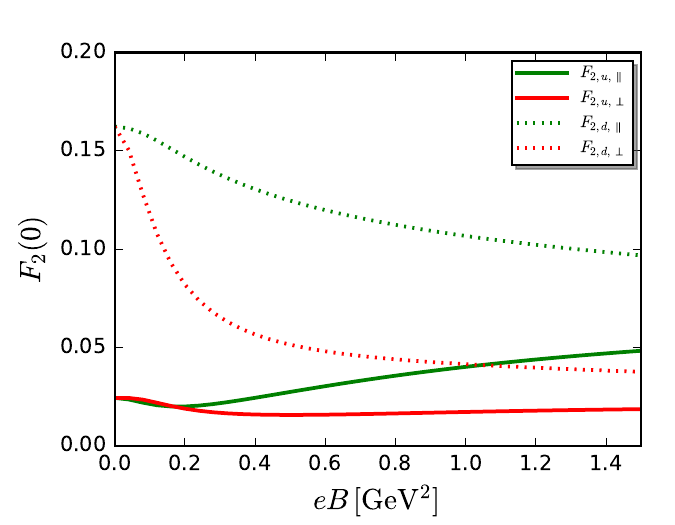}
\caption{The form factors as functions of the magnetic field strength. “Parallel” (longitudinal) or “perpendicular” (transverse) refers to the angle between the direction in which the photon's external momentum vanishes and the external magnetic field.}\label{fig:F2}
\end{figure}

\begin{figure}[t]
\includegraphics[width=0.5\textwidth]{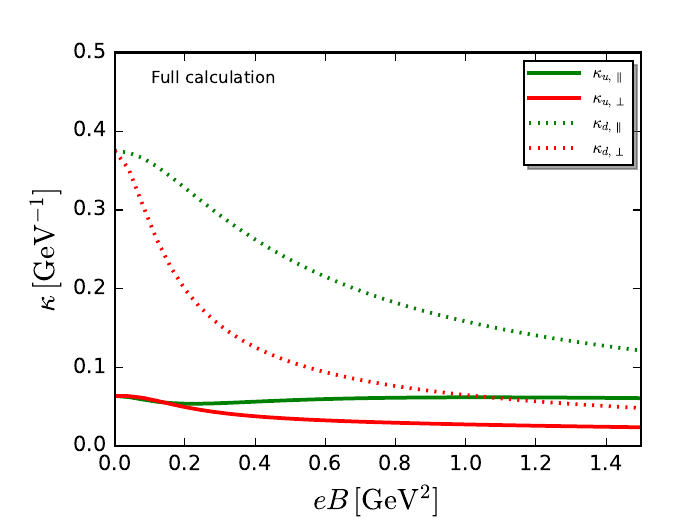}
\caption{The anomalous magnetic moments as functions of the magnetic field strength.}\label{fig:kappa}
\end{figure}

\begin{figure}[t]
\includegraphics[width=0.5\textwidth]{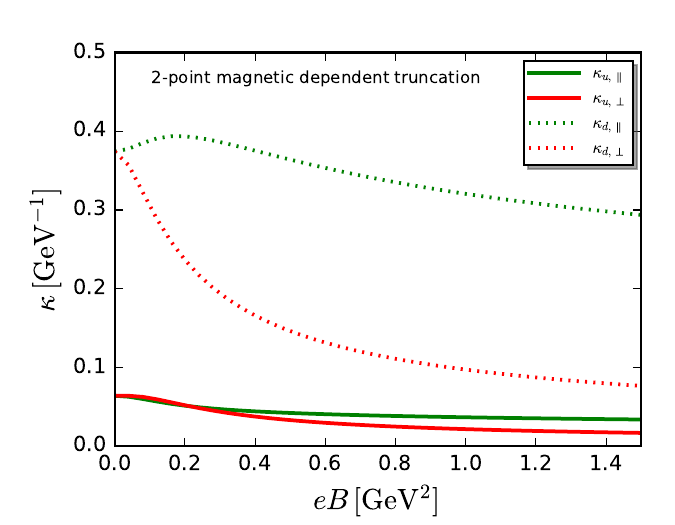}
\caption{The anomalous magnetic moments as functions of the magnetic field strength with 2-point magnetic dependent truncation. }\label{fig:kappa_2p}
\end{figure}

\begin{figure}[t]
\includegraphics[width=0.5\textwidth]{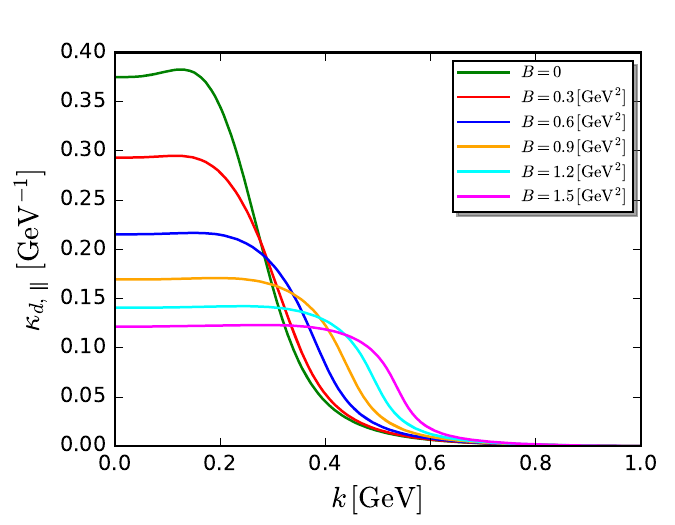}
\caption{The anomalous magnetic moment $\kappa_{d,\parallel}$ as functions of RG scale $k$ for several values of the magnetic field strength.}\label{fig:amm_k}
\end{figure}

We start the evolution of the FRG flows at an ultraviolet cutoff scale $\Lambda=1~\mathrm{GeV}$, and set the initial conditions $m_\Lambda=13.5~\mathrm{MeV}$, $\lambda_{4q,\Lambda}^{\pi}=\lambda_{4q,\Lambda}^{\sigma}=10.6\Lambda^{-2}$ and $\lambda_{4q,\Lambda}^{\alpha\notin\{\pi,\sigma\}}=0$. The quark mass in the vacuum is $m(T=0,B=0)=350~\mathrm{MeV}$. The pole mass of the pion is $m_\pi=141~\mathrm{MeV}$, which is determined by analytic continuation from the Euclidean to Minkowski spacetime using Pad\'{e} approximants. For more details, see \cite{Fu:2022uow,Fu:2024ysj,Fu:2025hcm}.

In the left panel of \Cref{fig:mf}, we present the quark masses as functions of RG scale $k$ for several values of the strength of magnetic field. The quark masses, which reflect chiral symmetry breaking, increase with the magnetic field. The chiral symmetry-breaking scales also increase with the magnetic field strength. These features reflect the magnetic catalysis effect of our theory at zero temperature. In the right panel of \Cref{fig:mf}, we show the quark masses as functions of the strength of magnetic field strength. The lattice QCD results are constructed from the quark chiral condensates in \cite{Ding:2020hxw}. We also compare our full calculations, in which both two- and four-point quark correlation functions are computed using the magnetic dependent quark propagators, with those from a simplified truncation scheme. In the simplified truncation scheme, the magnetic-dependent quark propagators are only used in the calculation of the flow equations of two-point correlation functions, while free quark propagators are employed in calculation of the flow equations of four-point quark correlation functions. Both truncation results exhibit the magnetic catalysis behavior. However, the full calculation shows a stronger dependence on the magnetic field strength and agrees better with the lattice QCD results. This indicates that an accurate treatment of the magnetic field dependence in the four-point quark correlation functions is important.

In \Cref{fig:lam}, we plot the four-quark couplings as functions of the strength of the magnetic field. Obviously, the $\pi$ and $\sigma$ channels are dominant, and $\lambda_{4q}^\pi > \lambda_{4q}^\sigma$ at $B=0$, since pions are pseudo-Goldstone bosons. As mentioned in \Cref{sec:2-flavorLEFT}, $\lambda_{4q}^\pi$ splits into $\lambda_{4q}^{\pi^0}$ and $\lambda_{4q}^{\pi^\pm}$ under magnetic fields. In this work, the splitting of $\lambda_{4q}^\pi$ is not taken into account, and the presented results can be regarded as an averaged approximation of these couplings. The couplings of $\pi$ and $\sigma$ channels rapidly decrease with the strength of the magnetic field in the small magnetic field region, and tend to become constant at large magnetic fields. A comparison of the dominant four-quark couplings as functions of the strength of magnetic field between full calculation and 2-piont magnetic dependent truncation is shown in \Cref{fig:lamcompare}. The magnetic dependence of the four-quark couplings in the two-point magnetic-field-dependent truncation comes from the quark mass, and the couplings decrease more slowly than those in the full calculation. In the right panel of \Cref{fig:lam}, we show the non-dominant four-quark couplings of the Fierz complete channels. We find that these non-dominant couplings contribute approximately $6\%$ to the quark masses and quark AMMs. The absolute values of these couplings also decrease with the strength of the magnetic field at $eB>0.2 ~\mathrm{GeV}^2$.

The dimensionless form factors as functions of the magnetic field strength are plotted in \Cref{fig:F2}, and the quark anomalous magnetic moments are plotted in \Cref{fig:kappa}. The magnitude of the AMM of the down quark is around 4 times larger than that of the up quark, as analyzed in \Cref{sec:AMM}. The transverse form factors and quark AMMs monotonically decrease with the magnetic field strength, while  the longitudinal u-quark AMM and form factor slightly increase with the magnetic field strength. To gain a deep understanding of these behaviors, we also plot the quark AMMs with the 2-piont magnetic dependent truncation (corresponding to the orange dashed line in \Cref{fig:mf}) in \Cref{fig:kappa_2p}. We expand the loop function \cref{eq:loop_kappa1} in a Taylor series around $B=0$ and ignore the magnetic dependence of the regulators:
\begin{align}
\mathscr{L}_{\kappa,\parallel}(f)=&\int \frac{q^3 dq}{(2 \pi)^2}\tilde{\partial}_t  \bigg\{  -\frac{m (r_q+1)}{2 \left(m^2+q^2 (r_q+1)^2\right)^2}\nonumber \\
&+\frac{m (r_q+1) \left(m^2-5 q^2 (r_q+1)^2\right)}{6 \left(m^2+q^2 (r_q+1)^2\right)^5} (q_f B)^2 \nonumber \\
&+ \mathscr{O}(q_f B)^4\bigg\}.
\end{align}
This expression can also be obtained from the weak-field expansion of the quark propagators \cite{Chyi:1999fc,  Wen:2023qcz}. The next-to-leading order term of longitudinal AMM is predominantly negative, which enhances the AMM. For the transverse loop function \cref{eq:loop_kappa2}:
\begin{align}
\mathscr{L}_{\kappa,\perp}(f)=&\int \frac{q^3 dq}{(2 \pi)^2}\tilde{\partial}_t  \bigg\{   -\frac{m (r_q+1)}{2 \left(m^2+q^2 (r_q+1)^2\right)^2}\nonumber \\
&+\frac{m^3 (r_q+1)}{\left(m^2+q^2 (r_q+1)^2\right)^5}(q_f B)^2 \nonumber \\
&+ \mathscr{O}(q_f B)^4\bigg\}.
\end{align}
The sign of the next-to-leading order term is opposite to that of the leading-order term, leading to a reduction in the transverse AMM. This analysis is consistent with the results in the small magnetic field region shown in \Cref{fig:kappa_2p}. However, in the full calculation, both longitudinal and transverse AMMs are suppressed due to the rapidly decreasing four-quark couplings, see \Cref{fig:lam} and \ref{fig:lamcompare}.

The transverse quark AMMs influence the dispersion relations of quarks under magnetic fields in other model studies, and a feedback calculation is left for future work. The transverse quark AMMs obtained in our calculation, which predominantly decrease with magnetic field strength, could avoid the unphysical ``jumps" in the case of constant $\kappa$ in \cite{Xu:2020yag}.

In \Cref{fig:amm_k}, we take $\kappa_{d,\parallel}$ as an example and plot it as a function for scale $k$ with several values of magnetic field strength of . The energy scale $k$ at which the quark AMMs are generated, decreases with the magnetic field strength. By comparing with the quark mass results in \Cref{fig:mf}, We find that the quark AMM generation energy scale is the same as the chiral symmetry breaking energy scale. This is easy to understand. As explained in \cite{Fu:2022uow}, the dominant channels of $\lambda_{4q}^{(\alpha)}$, i.e. $\lambda_{4q}^{\pi}$ and $\lambda_{4q}^{\sigma}$, rapidly increase at the energy scale of chiral symmetry breaking. The quark AMMs are proportional to the four-quark couplings $\lambda_{4q}^{(\alpha)}$, see \cref{eq:quark-photon}. As a result, the quark AMMs are generated with the chiral symmetry breaking.

In vacuum, i.e., at $B=0$, we also compute the proton and neutron magnetic moments using the constituent quark model (see \cref{eq:magneticmoments1}-(\ref{eq:magneticmoments3})). We obtain
\begin{align}
\mu_\text{proton}/\mu_N&=2.787 \\
\mu_\text{neutron}/\mu_N&=-1.995.
\end{align}
These results are close to the experimental results, i.e. $\mu_\text{proton}/\mu_N=2.7928$ and $\mu_\text{neutron}/\mu_N=-1.9130$ \cite{Mohr:2024kco}. 

\section{Summary and outlook}
\label{sec:summary}

We build a 2-flavor Nambu-Jona-Lasinio-type effective theory with four-quark scatterings under external magnetic fields within the FRG approach, in which Fierz-complete four-quark scatterings are self-consistently included through the RG flows. The quark mass increases with the strength of the magnetic field, reflecting the magnetic catalysis property. The quark mass as functions of the strength of magnetic field are in good agreement with the lattice QCD results. The $\pi$ and $\sigma$ channels are dominant, and their couplings decrease rapidly with the strength of magnetic field.

Based on this low-energy effective theory, we calculate quark AMMs and form factors.  By comparing the RG flows of the quark mass and AMMs, we find that the quark AMMs are dynamically generated with chiral symmetry breaking. The magnitude of the AMM ofthe  down quark is around 4 times larger than that of the up quark. The transverse quark AMMs monotonically decrease with magnetic field strength. The longitudinal d-quark AMM also monotonically decreases with the magnetic field strength due to the rapid decrease in the four-quark couplings, and the longitudinal u-quark AMM slightly increases with the magnetic field strength. At $B=0$, the magnetic moments of proton and neutron are computed using the constituent quark model, which are close to the experimental values.

Our quark AMMs and four-quark couplings results could serve as inputs for calculations in other effective models. We will extend our calculations to finite temperatures and chemical potential after considering the momentum dependence of the correlation functions. The first-principles QCD calculation of quark AMMs based on \cite{Fu:2025hcm} will be carried out in the upcoming works. In addition, a lattice QCD calculation of the quark AMMs could provide valuable guidance and a benchmark for further investigation.

\begin{acknowledgments}
We thank Jie Mei, Fan Lin and Shi-jun Mao for their valuable discussions. 
This work is supported in part by the National Natural Science Foundation of China (NSFC) Grant Nos: 12235016, 12221005, 12175030 and the Strategic Priority Research Program of Chinese Academy of Sciences under Grant No XDB34030000, the start-up funding from University of Chinese Academy of Sciences(UCAS), and the Fundamental Research Funds for the Central Universities. This work is also supported by the Collaborative Research Centre SFB 1225 (ISOQUANT).
\end{acknowledgments}

\bibliography{ref-lib}
	
\end{document}